\documentstyle[12pt]{article}

\begin{document}
\title{\bf {Casimir stress on parallel plates in de Sitter space }}
\author{
M.R. Setare $^1$ \footnote{E-mail: Mreza@physics.sharif.ac.ir}
\\  R. Mansouri $^{1,2}$\footnote{E-mail: mansouri@sina.sharif.ac.ir}  \\
 {$^1$ Department of Physics, Sharif University of
Technology, Tehran, Iran}\\ and \\ {$^{1,2}$ Institute for
Theoretical Physics and Mathematics, Tehran, Iran}}
\date{\small{\today}}
 \maketitle

\begin{abstract}
The Casimir stress on two parallel plates in de Sitter background
for massless scalar field satisfying Robin boundary conditions on
the plates is calculated. The metric is written in conformally
flat form to make maximum use of the Minkowski space
calculations. Different cosmological constants are assumed for
the space between and outside of the plates to have general
results applicable to the case of domain wall formations in the
early universe.
 \end{abstract}
% \begin{document}
\newpage
% \vspace*{10mm}

 \section{Introduction}
   The Casimir effect is one of the most interesting manifestations
  of nontrivial properties of the vacuum state in quantum field
  theory [1,2]. Since its first prediction by
  Casimir in 1948\cite{Casimir} this effect has been investigated for
  different fields having different boundary geometries[4-7]. The
  Casimir effect can be viewed as the polarization of
  vacuum by boundary conditions or geometry. Therefore, vacuum
  polarization induced by a gravitational field is also considered as
  Casimir effect.\\
  In the context of hot big bang cosmology, the unified theories
  of the fundamental interactions predict that the universe passes
  through a sequence of phase transitions. These phase transitions
  can give rise to domain structures determined by the topology of
  the manifold $M$ of degenerate vacuua \cite{{zel},{kib},{viel}}. If
  $M$ is disconnected, i.e. if $\pi(M)$ is nontrivial, then one
  can pass from one ordered phase to the other only by going
  through a domain wall. If $M$ has two connected components, e.g.
  if there is only a discrete reflection symmetry with
  $\pi_{0}(M)=Z_{2}$, then there will be just two ordered phase
  separated by a domain wall. In the domain wall formation models, in the
   early universe, the space-time changes from de Sitter to the
   geometry induced by the presence of a domain wall. In \cite{And}
   the effects of particle production and vacuum polarization attendant
   to the domain wall formation have been studied. Casimir stress for parallel
   plates in the background of static domain wall in four and two dimensions
  is calculated in \cite{{set1},{set2}}. Spherical bubbles immersed in different de Sitter
  spaces in- and out-side is calculated in \cite{set3}.  \\
  Our aim is to calculate the Casimir stress on two parallel plates with
  constant comoving distance having different vacuums between and outside, i.e. with
   false/true vacuum  between/outside. Our model may be used to study the effect of
   the Casimir force on the dynamics of the domain wall formation appearing in
  the simplest Goldston model. In this model potential of the scalar field
  has two equal minima corresponding to degenerate vacuua. Therefore, scalar field
  maps points at spatial infinity in physical space nontrivially into the
  vacuum manifold \cite{vil1}. Domain walls occur at the boundary between
  these regions of space. One may assume that the outer regions of parallel plates are
  in $\Lambda_{out}$ vacuum corresponding to degenerate vacuua in domain
  wall configuration. In section two we calculate the stress on
  two parallel plates with Robin boundary conditions. The case of
  different de Sitter vacuua between and outside of the plates, is
  considered in section three. The last section conclude and
  summarize the results.

\section{Parallel Plates with Robin boundary conditions in de Sitter Space }

Consider a massless scalar field coupled conformally to a de
Sitter background space. The scalar field satisfies the following
Robin boundary condition on two parallel plates within an
arbitrary space-time is defined as \cite{sah}:
 \begin{equation}
 (1+\beta_{m}(-1)^{m-1}
 \partial_{x})\Phi(x^{\nu})|_{x = a_{m}} = 0, \hspace{2cm}  m = 1,2,
  \end{equation}
Here we have assumed that the two plates are normal to the
cartesian $x$-axis at $x = a_{1, 2}$.\\
 The Robin boundary condition may be interpreted as the boundary condition on a thick
plate \cite{leb}. Rewriting(1) in the following form
\begin{equation}
\partial_{x}\Phi(x^{\nu})=(-1)^{m}\frac{1}{\beta_{m}}\Phi(x^{\nu}),
\end{equation}
where $|\beta_{m}|$, having the dimension of a length, may be
called skin-depth parameter. This is similar to the case of
penetration of an electromagnetic field into a real metal, where
the tangential component of the electric
field is proportional to the skin-depth parameter.\\
It is known that in the Minkowski space-time for the conformally
coupled scalar field the perpendicular pressure, $P$, is uniform
in the region between the plates and is given by \cite{sah}
\begin{equation}
 P=3\varepsilon_{c},
\end{equation}
where $\varepsilon_{c}$ is the Casimir energy density. This
Casimir energy has been calculated to be
\begin{equation}
\varepsilon = \varepsilon_{c}=
\frac{-A}{8\Gamma(5/2){\pi}^{3/2}a^{4}},
\end{equation}
where $A$ depends on $\beta_{1, 2}$ and may be inferred from the
Eq.(4.15) from the reference \cite{sah}. It has also been shown
that for $\beta_{1} = -\beta_{2}$
\begin{equation}
\varepsilon=\varepsilon_{c}=
\frac{-\zeta_{R}(4)\Gamma(2)}{(4\pi)^{2}a^{4}}=\frac{-\pi^{2}}{1440a^{4}},
\end{equation}
which is the same as for the Dirichlet and Neumann boundary
conditions.\\
 Consider now two parallel plates in the de Sitter
space-time. To make the maximum use of the flat space calculation
we use the de Sitter metric in the conformally flat form:
\begin{equation}
ds^{2}=\frac{\alpha^{2}}{\eta^{2}}
[d\eta^{2}-\sum_{\imath=1}^{3}(dx^{\imath})^{2}],
\end{equation}
where $\eta$ is the conformal time:
\begin{equation}
\infty < \eta < 0.
\end{equation}
The relation between parameter $\alpha$ and cosmological constant
$\Lambda$ is given by
\begin{equation}
\alpha^{2}=\frac{3}{\Lambda}.
\end{equation}
Using the standard relation between the energy-momentum tensor
for conformally coupled situations \cite{davies}
\begin{equation}
<T^{\mu}_{\nu}[\tilde
g_{\alpha\beta}]>=(\frac{g}{\tilde{g}})^{\frac{1}{2}}
<T^{\mu}_{\nu}[{g^{(M)}_{\alpha\beta}}]>-\frac{1}{2880}[\frac{1}{6}
\tilde H^{(1)\mu}_{\nu}-\tilde H^{(3)\mu}_{\nu}],
\end{equation}
where $g_{\mu \nu}$ and $\tilde g _{\mu \nu}$ are conformal to
each other, with their respective determinants $g$ and
$\tilde{g}$. We are going to assume now that $g_{\mu \nu}$ is the
Minkowski metric. Now,
$<T^{\mu}_{\nu}[{g^{(M)}_{\alpha\beta}}]>$, the regulatized
energy momentum tensor for a conformally coupled scalar field for
the case of parallel plate configuration in flat space-time is
given by
\begin{equation}
<T^{\mu}_{\nu}[{g^{(M)}_{\alpha\beta}}]>=diag(\varepsilon,-P,-P_{\bot},-P_{\bot})
=diag(\varepsilon,3\varepsilon,-\varepsilon,-\varepsilon).
\end{equation}
The second term in (9)is the vacuum polarization due to the
gravitational field, without any boundary conditions. The
functions $H^{(1,3)\mu}_{\nu}$ are some combinations of curvature
tensor components (see \cite{davies}). For massless scalar field
in de Sitter space, the term is given by \cite{{davies},{Dowk}}
\begin{equation}
-\frac{1}{2880}[\frac{1}{6} \tilde H^{(1)\mu}_{\nu}-\tilde
H^{(3)\mu}_{\nu}]
=\frac{1}{960\pi^{2}\alpha^{4}}\delta^{\mu}_{\nu}.
\end{equation}
From(3,5,8,9,10) one can obtain vacuum pressure due to the
boundary acting on the plates:
\begin{equation}
P^{(1,2)}_{b}=P_{b}(x_{1,2})=(\frac{g}{\tilde{g}})^{\frac{1}{2}}
(\frac{-3\pi^{2}}{1440 a^{4}})=\frac{\eta^{4}}{\alpha^{4}}
(\frac{-3\pi^{2}}{1440 a^{4}})=\frac{-\eta^{4}\Lambda^{2}}{3}
\frac{\pi^{2}}{1440 a^{4}},
\end{equation}
which is attractive. It has been shown that this pressure is zero
for $x < a_1$ and $x > a_2$ \cite{{set1}, {set2}}. The
gravitational part of the pressure according to (11) is equal to
\begin{equation}
P_{g}=-<T^{1}_{1}>=\frac{-1}{960\pi^{2}\alpha^{4}}.
\end{equation}
This is the same from both sides of the plates, and hence leads
to zero effective force. Therefore the effective force acting on
the plates are given only by the boundary part.

\section{Parallel plates with different cosmological constants
 between and out-side}
 Now, assume there are different vacuua between and out-side of
 the plates, corresponding to $\alpha_{betw}$ and $\alpha_{out}$ in the
 metric(6). As we have seen in the last section, the vacuum
 pressure due to the boundary is only non-vanishing between the
 plates. Therefore, we have for the pressure due to the boundary
 \begin{equation}
 P^{(1,2)}_{b}=\frac{\eta^{4}}{\alpha_{betw}^{4}}\frac{-3\pi^{2}}{1440a^{4}}
 =\frac{-\eta^{4}\Lambda_{betw}^{2}}{3}\frac{\pi^{2}}{1440a^{4}}.
 \end{equation}
 Now, the effective pressure created by gravitational part(11), is
 different for different part of the space-time:
 \begin{equation}
 P^{betw}_{g}=-<T^{1}_{1}>_{betw}=\frac{-1}{960\pi^{2}\alpha_{betw}^{4}}=
 \frac{-\Lambda_{betw}^{2}}{9}\frac{1}{960\pi^{2}},
 \end{equation}
 \begin{equation}
 P^{out}_{g}=-<T^{1}_{1}>_{out}=\frac{-1}{960\pi^{2}\alpha_{out}^{4}}=
 \frac{-\Lambda_{out}^{2}}{9}\frac{1}{960\pi^{2}}.
 \end{equation}
 Therefore, the gravitational pressure acting on the plates is
 given by
 \begin{equation}
 P_{g}=P^{betw}_{g}-P^{out}_{g}=\frac{-1}{9\times 960\pi^{2}}
 (\Lambda_{betw}^{2}-\Lambda_{out}^{2}).
 \end{equation}
 The total pressure acting on the plates, $P$, is then given by
 \begin{equation}
 P=P_{g}+P_{b}=\frac{-1}{9\times 960\pi^{2}}
 (\Lambda_{betw}^{2}-\Lambda_{out}^{2})-
 \frac{\eta^{4}\Lambda_{betw}^{2}}{3}\frac{\pi^{2}}{1440a^{4}}.
 \end{equation}
 The term $P_{b}$ is always negative corresponding to an
 attractive force on the plates. The term $P_{g}$, however, may
 be negative or positive, depending on the difference between the
 cosmological constants in the two parts of space-time. Given a false
 vacuum between the plates, and true vacuum out-side, i.e. $\Lambda_{betw}>
 \Lambda_{out}$, then the gravitational part is negative.
 Therefore, the total pressure $P$ is always negative leading to a
 attraction of the plates. For the case of true vacuum between
 the plates and false vacuum out-side, i.e.
 $\Lambda_{betw}<\Lambda_{out}$, the gravitational pressure is
 positive. Therefore, the total pressure may be either negative or
 positive. For $P>0$, the initial repulsion of the parallel plates
 may be stopped or not depending on the detail of the
 dynamics. Given $P<0$ initially, it remains negative and there is an attraction between
 the plates.

 \section{Conclusion}

 We have considered two parallel
 plates in de Sitter background with a massless scalar field,
 coupled conformally to it, satisfying the Robin
 boundary conditions with constant comoving distance. Our calculation
 shows that for the parallel plates with false vacuum between and true
  vacuum outside, the total Casimir force leads to an attraction of the
  plates. The boundary term is proportional to the forth power of the inverse
 distance between the plates, and is always negative, which means a huge
 attractive force for small distances. Therefore, parallel plates with false
 vacuum in between always attract each other. In contrast, plates with true
 vacuum between them may repel each other to a maximum distance
 and attract again. The result may be of interest in the case of
 formation of the cosmic domain walls in early universe, where the
  wall orthogonal to the $x-$axis is described by the
 function $\Phi_{i}(x)$ interpolating between two different minima
 at $x\rightarrow \pm\infty$ \cite{vil1}.

  \vspace{3mm}

{\large {\bf  Acknowledgement }}\\
 We would like to thank Prof. A. A. Saharian for his
 valuabel hints and comments.
 \vspace{1mm}
\small.

\end{document}